\documentclass[aps,pra,twocolumn,superscriptaddress]{revtex4-2}
\usepackage{graphicx}
\usepackage{amsmath}
\usepackage{amsthm}

\usepackage{amsbsy}
\usepackage{array}
\usepackage{amssymb}
\usepackage{multirow}
\usepackage{verbatim}
\usepackage{caption} 
\usepackage{subcaption} 
\usepackage{float} 
\usepackage{hyperref}
\usepackage{epstopdf}
\usepackage{footnote}
\usepackage{lipsum} 

\hypersetup{
	colorlinks=true,  
	linkcolor=blue,   
	citecolor=blue,   
	urlcolor=blue,    
	pdfborder={0 0 0}, 
}

\begin{document}
\title{Semi-device-independent certification of high-dimensional quantum channels}

\author{Mengyan Li}
\affiliation{School of Mathematical Sciences, Beijing University of Posts and Telecommunications, Beijing 100876, China}
\affiliation{Key Laboratory of Mathematics and Information Networks, Beijing University of Posts and Telecommunications, Ministry of Education, Beijing 100876, China}
\affiliation{State Key Laboratory of Networking and Switching Technology, Beijing University of Posts and Telecommunications, Beijing 100876, China}

\author{Yanning Jia}
\affiliation{School of Mathematical Sciences, Beijing University of Posts and Telecommunications, Beijing 100876, China}
\affiliation{Key Laboratory of Mathematics and Information Networks, Beijing University of Posts and Telecommunications, Ministry of Education, Beijing 100876, China}
\affiliation{State Key Laboratory of Networking and Switching Technology, Beijing University of Posts and Telecommunications, Beijing 100876, China}

\author{Fenzhuo Guo}\email{gfenzhuo@bupt.edu.cn}
\affiliation{School of Mathematical Sciences, Beijing University of Posts and Telecommunications, Beijing 100876, China}
\affiliation{Key Laboratory of Mathematics and Information Networks, Beijing University of Posts and Telecommunications, Ministry of Education, Beijing 100876, China}
\affiliation{State Key Laboratory of Networking and Switching Technology, Beijing University of Posts and Telecommunications, Beijing 100876, China}

\author{Haifeng Dong}
\affiliation{School of Instrumentation Science and Opto-Electronics Engineering, Beihang University, Beijing 100191, China}

\author{Sujuan Qin}
\affiliation{State Key Laboratory of Networking and Switching Technology, Beijing University of Posts and Telecommunications, Beijing 100876, China}

\author{Fei Gao}
\affiliation{State Key Laboratory of Networking and Switching Technology, Beijing University of Posts and Telecommunications, Beijing 100876, China}

\begin{abstract}
	
	Certifying high-dimensional quantum channels is essential for ensuring the reliability of quantum communication protocols. Existing certification schemes often rely on fully trusted internal devices, which is difficult to achieve in realistic scenarios. Here, we propose a semi-device-independent framework for certifying channel properties directly from observed statistics, assuming only that the system dimension is known. By explicitly incorporating the full set of structural constraints inherent to Choi states, our approach exploits the Choi-Jamiołkowski isomorphism for rigorous certification of quantum channels. The entanglement dimensionality of quantum channels is first certified by introducing a witness and numerically determining its Schmidt-number-dependent bounds. This certification method reproduces known analytical benchmarks and is applied to dephasing and depolarizing noise channels, thereby confirming its validity. To provide a more complete assessment of channel performance, the entanglement fidelity of quantum channels is also certified using a hierarchy of semidefinite programming relaxations based on localizing matrices. Lower bounds on the entanglement fidelity are obtained that are compatible with either the full set of observed statistics or a single witness value.
		
\end{abstract}
\maketitle

\section{\label{sec1:level1}INTRODUCTION}

Quantum communication is a central pillar of quantum information science, enabling fundamentally new modes of information transfer that surpass classical limits in security, efficiency, and functionality \cite{Gisin:2002zz,2008kimbleQuantumInternet,Wehner:2018enf}. In practice, quantum communication tasks are implemented through specific protocols, each relying on the physical process governing the transmission of quantum states from a sender to a receiver, typically in the presence of noise and environmental interactions. Such processes are mathematically described by quantum channels \cite{nielsenQuantumComputationQuantum2010}, which represent the most general evolutions permitted by quantum mechanics for open quantum systems. Understanding and assessing these channels is therefore essential for evaluating the performance and fundamental limits of quantum communication protocols.

Traditionally, quantum channels have been characterized using quantum process tomography (QPT), which aims to fully reconstruct the channel’s process matrix from tomographic data \cite{Poyatos:1996mf,2001darianoQuantumTomography,2008mohseniQuantumprocessTomography}. Although powerful in principle, QPT suffers from exponential resource scaling with system dimension and requires precise control and trust in all internal devices. To overcome these limitations, measurement-device-independent (MDI) and device-independent (DI) certification schemes have been developed. In MDI schemes, quantum channels are certified while relaxing assumptions about the measurement devices \cite{Pusey:2015jln,rossetResourceTheoryQuantum2018,abiusoVerificationContinuousvariableQuantum2023}, whereas in DI schemes, channel properties are inferred solely from Bell-like correlations, removing assumptions about all internal device structures \cite{dallarnoDeviceindependentTestsQuantum2017,sekatskiCertifyingBuildingBlocks2018,wagnerDeviceindependentCharacterizationQuantum2020,chenRobustSelftestingSteerable2021,Sekatski:2023ukn}. Experimental demonstrations on several physical platforms have confirmed the practical feasibility of these schemes \cite{maoExperimentallyVerifiedApproach2020,graffittiMeasurementDeviceIndependentVerificationQuantum2020,Yu:2021pxu,nevesExperimentallyCertifiedTransmission2025}. To date, the above MDI and DI schemes are mainly focused on qubit systems. 

High-dimensional quantum systems have advantages in enhancing communication capacity \cite{2018erhardTwistedPhotons,Cozzolino:2019gjf}, increasing noise resilience \cite{Ecker:2019uzb,zhuHighdimensionalPhotonicEntanglement2021}, and enabling more complex information processing tasks \cite{Valencia:2020aon}. Recent efforts have begun to extend channel certification schemes beyond qubit systems. In this direction, Mallick \emph{et al.} \cite{mallickHigherdimensionalentanglementDetectionQuantumchannel2025} introduced methods based on moments of generalized positive maps to certify the entanglement dimensionality, as quantified by Schmidt number \cite{terhalSchmidtNumberDensity2000}, of quantum channels. More recently, Ref.~\cite{engineerCertifyingHighdimensionalQuantum2025} proposed entanglement dimensionality witnesses to estimate the minimal entanglement dimensionality preserved during transmission. While both works establish an important foundation for high-dimensional channel certification via the Choi-Jamiołkowski (CJ) isomorphism \cite{jiangChannelstateDuality2013}, which maps a quantum channel to an associated bipartite state commonly referred to as the Choi state, they do not explicitly incorporate the partial-trace constraint intrinsic to Choi states. As a consequence, the certifiable entanglement dimensionality may not be stringent enough. Furthermore, these schemes rely on fully trusted internal devices, which limits their practical applicability in quantum communication tasks.

\begin{figure}
	\centering
	\includegraphics[width=5.8cm]{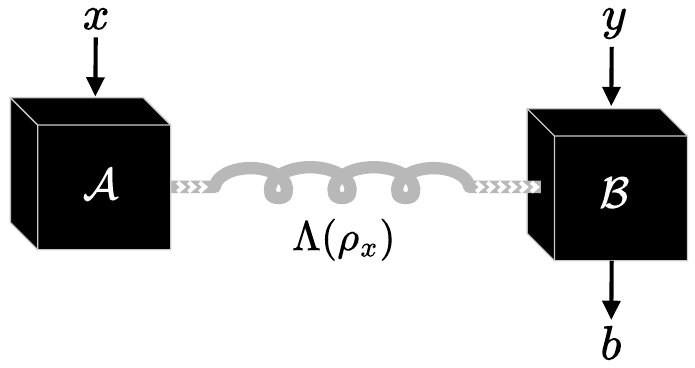}
	\caption{Diagram of the SDI-P$\&$M scenario. $\mathcal{A}$ uses an unspecified preparation device that emits a quantum state $\rho_x$ depending on the classical input $x$. This state is transmitted through a quantum channel $\Lambda$. $\mathcal{B}$ receives the transmitted state and performs a measurement using an unspecified measurement device, producing an output $b$ conditioned on the classical input $y$. In this scenario, only the system dimension $d$ is assumed to be known. }

	\label{fig1}
\end{figure}

In this work, we certify high-dimensional quantum channels in a semi-device-independent (SDI) scenario (see Fig.~\ref{fig1}), where only the system dimension $d$ is assumed to be known, while all other aspects of the preparation and measurement devices are left unspecified. Channel properties are inferred directly from the observed prepare-and-measure (P$\&$M) statistics, and the partial-trace constraint intrinsic to Choi states is explicitly enforced to ensure consistency with a valid quantum channel description. Within this framework, we first certify the entanglement dimensionality of quantum channels by introducing a witness inspired by the average success probability (ASP) of quantum random access codes (RACs)~\cite{tavakoliQuantumRandomAccess2015}. Through numerical optimization over Choi states that explicitly incorporate both structural and Schmidt number constraints, we derive Schmidt-number-dependent bounds for this witness, which recover known analytical benchmarks from Ref.~\cite{tavakoliQuantumRandomAccess2015} in specific cases. We further apply our method to dephasing and depolarizing noise channels, thereby establishing a direct relation between noise parameters and certifiable entanglement dimensionality. Nevertheless, entanglement dimensionality alone does not provide a complete description of channel performance. In particular, even when two channels share the same entanglement dimensionality, the strength of the entanglement they preserve may differ substantially. To capture this aspect, we go beyond entanglement dimensionality and develop a method for certifying entanglement fidelity~\cite{Horodecki:1998zh} directly from observed statistics. To this end, we construct a hierarchy of semidefinite programming (SDP) relaxations based on localizing matrices~\cite{tavakoliSemidefiniteProgrammingRelaxations2024}, which provide systematic lower bounds on the entanglement fidelity compatible with the full set of observed statistics. Finally, we show that the entanglement fidelity can also be certified by incorporating the observed witness value as a constraint within the SDP hierarchy.

\section{\label{sec2:level1}PRELIMINARIES}

We consider the SDI-P$\&$M scenario involving two parties, $\mathcal{A}$ and $\mathcal{B}$, who are connected only through a quantum channel. Generally, $\mathcal{A}$ prepares states denoted by $\rho_x$, where $x=(x_1,\ldots,x_n)$ with $x_i\in[d]:=\{1,2,\ldots,d\}$. $\mathcal{B}$ performs measurements described by positive operator valued measure (POVM) elements $M_{b|y}$, with $y\in[n]$ and $b\in[d]$, satisfying $M_{b|y}\succeq 0$ and $\sum_{b} M_{b|y}=\mathbb{I}_\mathcal{B}$. The conditional probability distribution observed in the experiment is therefore
\begin{equation}\label{equ1}
	P(b|x,y)=\operatorname{Tr}\!\left[\Lambda(\rho_x)M_{b|y}\right],
\end{equation}
where $\Lambda$ denotes the quantum channel from $\mathcal{A}$ to $\mathcal{B}$. Let $\mathcal{M}_d$ represent the set of complex matrices of dimension $d \times d$. A quantum channel is a linear map $\Lambda:\mathcal{M}_d \mapsto \mathcal{M}_d$ that is completely positive and trace preserving (CPTP) \cite{nielsenQuantumComputationQuantum2010}. Complete positivity requires that $(\mathrm{id}_k\otimes \Lambda)(X)\succeq 0$ for all $k\in\mathbb{N}$ and all positive semidefinite operators $X$, where $\mathrm{id}_k$ denotes the identity map acting on $\mathcal{M}_k$. This condition guarantees that $\Lambda$ preserves positivity when acting on a subsystem of an arbitrary larger composite system. Trace preservation further requires that $\operatorname{Tr}[\Lambda(\rho)] = \operatorname{Tr}[\rho]$ for all density operators $\rho$.

For the certification of quantum channels, it is often convenient to employ the CJ  isomorphism~\cite{jiangChannelstateDuality2013}, which maps a channel $\Lambda$ to its associated Choi state
\begin{equation}\label{equ2}
	\Phi_\Lambda = (\mathrm{id}\otimes \Lambda)\!\left(|\Phi^+_d\rangle\langle\Phi^+_d|\right),
\end{equation}
where $|\Phi^+_d\rangle=(1/\sqrt{d})\sum_{i=1}^{d}|i,i\rangle$ is a $d$-dimensional maximally entangled state. The map $\Lambda\mapsto\Phi_\Lambda$ is linear and injective \cite{jiangChannelstateDuality2013}, and its image is fully characterized by simple structural constraints. First, $\Phi_\Lambda\succeq 0$ because $\Lambda$ is completely positive; second, $\operatorname{Tr}(\Phi_\Lambda)=1$; and third, since $\Lambda$ acts on the second subsystem of the maximally entangled state, the reduced state on subsystem $\mathcal{A}$ satisfies
\begin{equation}\label{equ3}
	\operatorname{Tr}_\mathcal{B}(\Phi_\Lambda)=\frac{1}{d}\mathbb{I}_\mathcal{A}.
\end{equation}
The Choi representation in Eq.~\eqref{equ2} furnishes a complete parametrization of quantum channels, reducing the study of such channels to that of bipartite states subject to a fixed partial-trace constraint. We now rewrite the conditional probabilities in Eq.~\eqref{equ1} in a form that makes the role of the Choi state explicit. Using the adjoint map $\Lambda^\dagger$ and standard identities involving maximally entangled states, one obtains
\begin{equation}\label{equ4}
	\begin{aligned}
		P(b|x,y)
		&= \mathrm{Tr}\!\left[\Lambda(\rho_x) M_{b|y}\right] \\
		&= \mathrm{Tr}\!\left[\rho_x \Lambda^\dagger(M_{b|y})\right] \\
		&= d\,\mathrm{Tr}\!\left[(\rho_x^{T}\!\otimes \Lambda^\dagger(M_{b|y}))|\Phi_d^{+}\rangle\langle\Phi_d^{+}|\right] \\
		&= d\,\mathrm{Tr}\!\left[(\rho_x^{T}\!\otimes M_{b|y})\Phi_\Lambda\right].
	\end{aligned}
\end{equation}

Two widely used figures of merit in high-dimensional quantum systems are entanglement dimensionality \cite{terhalSchmidtNumberDensity2000} and entanglement fidelity \cite{Horodecki:1998zh}. The notion of entanglement dimensionality aims to capture the minimal local Hilbert space dimension required to reproduce a given bipartite quantum state. For pure states, this quantity is quantified by the Schmidt rank; for mixed states, an operational extension is provided by the Schmidt number \cite{terhalSchmidtNumberDensity2000}, defined as
\begin{equation}\label{equ5}
	\mathrm{SN}(\rho)=\min_{\{p_i,|\psi_i\rangle\}} \max_i \ \mathrm{SR}(|\psi_i\rangle),
\end{equation}
where the minimization runs over all convex decompositions $\rho=\sum_i p_i|\psi_i\rangle\langle\psi_i|$ and $\mathrm{SR}(|\psi_i\rangle)$ denotes the Schmidt rank of $|\psi_i\rangle$. In the context of quantum channels, the CJ isomorphism establishes that
$\mathrm{SN}(\Lambda)=\mathrm{SN}(\Phi_\Lambda)$. This equality implies that certifying the Schmidt number of a channel is operationally equivalent to certifying the Schmidt number of its Choi state. For convenience, we shall refer to this quantity as the entanglement dimensionality of quantum channels.

We note that two channels with the same entanglement dimensionality may nonetheless differ significantly in the amount of entanglement they preserve. To capture this perspective, we adopt the entanglement fidelity \cite{Horodecki:1998zh} as an additional figure of merit. Formally, the entanglement fidelity of a channel $\Lambda$ is defined as
\begin{equation}\label{equ6}
	\mathcal{F}(\Phi_\Lambda)=\max_{\Xi}\;
	\langle\Phi_d^{+}|(\mathrm{id}\otimes\Xi)[\Phi_\Lambda]|\Phi_d^{+}\rangle,
\end{equation}
where the maximization is over all local extraction channels $\Xi$ acting on the second subsystem. 

\section{\label{sec3:level1}Certifying the Entanglement Dimensionality of Quantum Channels}

The performance of quantum communication tasks is constrained by the extent to which a quantum channel can preserve entanglement. Based on this physical connection, we introduce a Schmidt number constraint $\mathrm{SN}(\Lambda)\le r$, with $r\in[d]$, and determine its impact on communication performance.

We now turn to a specific communication task: an $n$-input, $d$-level RAC~\cite{tavakoliQuantumRandomAccess2015}, with performance evaluated through the ASP
\begin{equation}\label{equ7}
	\alpha_{n,d}:=\frac{1}{nd^n}\sum_{x\in[d]^n}\sum_{y\in[n]}P[b=x_y|x,y].
\end{equation}
For a fixed Schmidt number bound $r$, we define the optimal achievable ASP in a quantum RAC as
\begin{equation}\label{equ8}
	\beta_{n,d,r}^{\mathcal{Q}}
	:=\max_{\{\rho_x\},\{M_{b|y}\}}
	\frac{1}{nd^n}\sum_{x,y}\mathrm{Tr}\!\left[\Lambda(\rho_x)M_{x_y|y}\right].
\end{equation}
When $r=1$, the channel is entanglement breaking~\cite{horodeckiEntanglementBreakingChannels2003} and therefore incapable of transmitting any entanglement. In this regime, the RAC protocol admits no genuinely quantum advantage, and the achievable ASP is necessarily bounded by the classical RAC limit. For intermediate values $1<r<d$, the channel is partially entanglement breaking~\cite{chruscinskiPartiallyEntanglementBreaking2006}, in the sense that entanglement can be preserved only up to a finite Schmidt number. This restricts the entanglement structure accessible at the decoding stage, which limits the attainable RAC performance. Finally, when $r=d$, the channel can, in principle, preserve full $d$-dimensional entanglement, allowing unrestricted quantum decoding and thereby enabling the optimal quantum RAC performance. By combining the equality in Eq.~\eqref{equ4}, the optimization problem in Eq.~\eqref{equ8} can be equivalently rewritten in the explicit form
\begin{equation}
	\begin{aligned}\label{equ9}
		\max_{\Phi_\Lambda,\{\rho_x\},\{M_{b|y}\}} \ \ &\frac{1}{nd^{n-1}} \sum_{x,y} \mathrm{Tr}\left[(\rho_x \otimes M_{x_y|y}) \Phi_\Lambda\right],\\
		\mathrm{s.t.~~~~~~~~} 
		& \Phi_\Lambda \in \mathcal{S}_r,\\
		&\mathrm{Tr}_\mathcal{B}(\Phi_\Lambda)=\mathbb{I}_\mathcal{A}/d, \\
		& \rho_x \succeq 0, \mathrm{Tr}(\rho_x) = 1, \\
		& M_{b|y}\succeq0, \sum_b M_{b|y}=\mathbb{I}_\mathcal{B}, 
	\end{aligned}
\end{equation}
where the transpose on $\rho_x$ is omitted for notational simplicity, which does not affect the results since the state preparation device is treated as unspecified. Moreover, $\mathcal{S}_r$ denotes the set of bipartite states with Schmidt number at most $r$, namely $\mathcal{S}_r := \{\sigma_{\mathcal{AB}} : \mathrm{SN}(\sigma_{\mathcal{AB}})\le r\}$. The ASP compatible with channels of Schmidt number at most~$r$ is upper bounded by $\beta_{n,d,r}^{\mathcal{Q}}$. Consequently, if the experimentally observed ASP violates this bound, i.e., $ \alpha_{n,d}>\beta_{n,d,r}^{\mathcal{Q}}$, then the underlying channel $\Lambda$ must necessarily satisfy $\mathrm{SN}(\Lambda)\ge r+1$.

Eq.~\eqref{equ9} involves a simultaneous optimization over the Choi state, state preparations, and measurements, which renders the problem intrinsically nonconvex. Furthermore, the partial-trace condition imposes a strong structural restriction on admissible Choi states. As a result, many existing Schmidt number certification methods developed for general bipartite states~\cite{bavarescoMeasurementsTwoBases2018,morelliResourceEfficientHighDimensionalEntanglement2023,Li:2024ewe} cannot be straightforwardly employed in our scenario. In addition, we note that the partial trace of $\Phi_\Lambda$ must be maximally mixed (rank $d$); consequently, $\Phi_\Lambda$ cannot be pure for any $\mathrm{SN}(\Phi_\Lambda)\le r<d$, as its marginal rank would be bounded by $r$. It follows that one cannot rely on pure state Schmidt decompositions to construct projectors that reduce the effective Hilbert space dimension from $d$ to $r$, and therefore some numerical techniques~\cite{pauwelsAlmostQuditsPrepareandmeasure2022,DAlessandro:2024qmh,2025rochicarcellerBoundEntangled} are not applicable to the present problem.

We adopt here an alternating convex search method~\cite{tavakoliSemidefiniteProgrammingRelaxations2024} to tackle the problem in Eq.~\eqref{equ9}. The method proceeds by iteratively optimizing over one subset of variables while keeping the remaining two fixed, cycling through the blocks $\{\rho_x\}$, $\{M_{b|y}\}$, and $\Phi_\Lambda$ until convergence; to mitigate convergence to local optima, we perform multiple random initializations and retain the best result. It is worth emphasizing that strictly imposing the constraint $\mathrm{SN}(\Phi_\Lambda)\le r$ is technically demanding~\cite{2020weilenmannEntanglementDetection}. We therefore employ two approximation techniques to render the problem tractable for numerical optimization. First, we adopt the generalized Doherty-Parrilo-Spedalieri (DPS) hierarchy introduced in Ref.~\cite{2020weilenmannEntanglementDetection}, which provides a systematic outer approximation to the set $\mathcal{S}_r$. Second, we apply the generalized reduction map (GRM)~\cite{terhalSchmidtNumberDensity2000} as a necessary criterion: any state in $\mathcal{S}_r$ must satisfy the corresponding operator inequality. The former offers higher characterization accuracy, although at comparatively higher computational cost even at low hierarchy levels, whereas the latter is extremely efficient but yields a coarser approximation. Further implementation details are provided in the Appendix. 

\begin{figure}
	\centering
	\includegraphics[width=8.7cm]{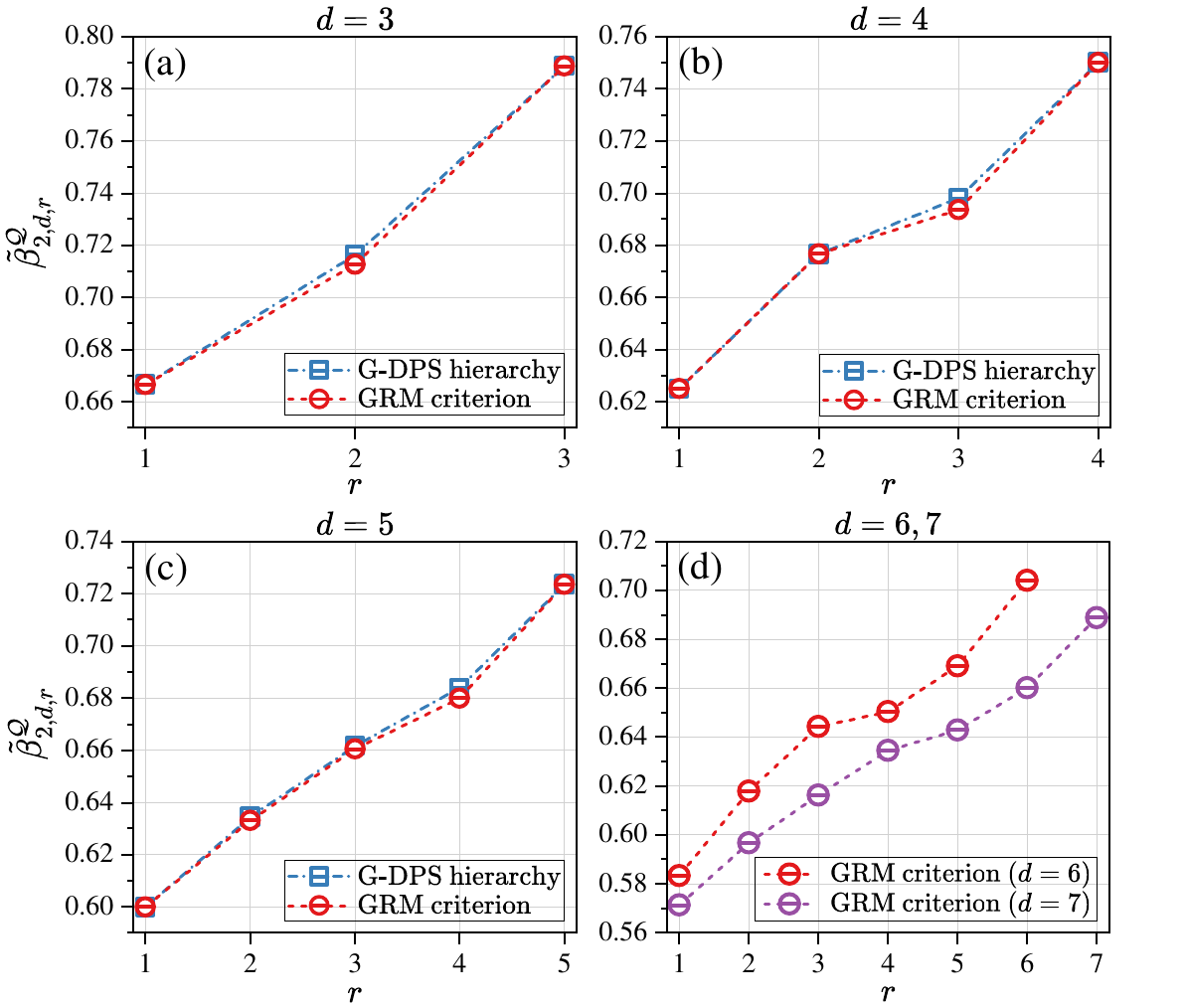}
	\caption{Numerical bounds $\tilde{\beta}_{2,d,r}^{\mathcal{Q}}$ for $d=3,\dots,7$. Panels (a)–(c) correspond to $d=3,4,5$, respectively. Blue squares denote the values computed by enforcing the Schmidt number constraint through the generalized DPS hierarchy (evaluated at the lowest level of the hierarchy), while red circles indicate the values obtained using the GRM criterion. Panel (d) shows the results for $d=6$ and $7$, where the Schmidt number constraint is imposed via the GRM criterion.
	}
	\label{fig2}
\end{figure}

We now present numerical results for the case $n=2$. It is known from Ref.~\cite{tavakoliQuantumRandomAccess2015} that the ASP of classical and quantum RACs satisfy
\begin{equation}\label{equ10}
	\centering
	\begin{aligned}
		\alpha_{2,d}^{\mathcal{L}} \le \frac{1}{2}\left(1+\frac{1}{d}\right):= \beta_{2,d}^{\mathcal{L}}, \\
		\alpha_{2,d}^{\mathcal{Q}} \le \frac{1}{2}\left(1+\frac{1}{\sqrt{d}}\right):= \beta_{2,d}^{\mathcal{Q}}.
	\end{aligned}
\end{equation}
Figs.~\hyperref[fig2]{2(a)–2(c)} display the numerical bounds $\tilde{\beta}_{2,d,r}^{\mathcal{Q}}$ for $d=3,4,5$ and $r=1,\ldots,d$, where the tilde denotes numerical results, and the same notation is used throughout. In the cases $r=1$ and $r=d$, the numerical bounds $\tilde{\beta}_{2,d,1}^{\mathcal{Q}}$ and $\tilde{\beta}_{2,d,d}^{\mathcal{Q}}$ coincide, within numerical precision ($\sim10^{-8}$), with the analytical benchmarks $\beta_{2,d}^{\mathcal{L}}$ and $\beta_{2,d}^{\mathcal{Q}}$, respectively. This agreement provides a nontrivial consistency check, confirming that the numerical optimization does not suffer from spurious local optima and that the Schmidt number constraints are correctly enforced. While $\tilde{\beta}_{2,d,r}^{\mathcal{Q}}$ represents a lower bound on $\beta_{2,d,r}^\mathcal{Q}$, comparison with the known benchmarks demonstrates that this bound is tight and reliable. Notably, the generalized DPS hierarchy and the GRM criterion produce nearly identical numerical results across all tested cases. This reflects the fact that, for the RAC instance considered here, both relaxations adequately capture the relevant Schmidt number constraints, resulting in only minor numerical deviations. Moreover, without the partial-trace constraint in Eq.~\eqref{equ9}, the resulting optimization becomes uninformative: the numerical bound $\tilde{\beta}_{2,d,r}^{\mathcal{Q}}$ attains the trivial value $1$ independently of $r$ and $d$. This demonstrates the necessity of the partial-trace condition for physically meaningful certification.

Although the generalized DPS hierarchy can in principle provide a tighter characterization of the set $\mathcal{S}_r$, its computational cost increases rapidly with the system dimension, even at the lowest hierarchy level. For $d>5$, the method becomes impractical due to memory and running-time limitations on standard computing resources (all numerical optimizations were performed on a PC with an Intel Core i9-12900K CPU, 32\textasciitilde GB RAM, using the MOSEK \cite{mosek} SDP solver). Given that the close numerical results are observed in the cases where both approaches are feasible, we therefore adopt the GRM criterion as the Schmidt number constraint for higher dimensions. Illustrative examples for $d=6$ and $7$ are shown in Fig.~\hyperref[fig2]{2(d)}.

Based on the numerical results reported above, we now discuss the practical relevance of our findings in noisy experimental settings. Realistic quantum channels are typically subject to decoherence or other imperfections, so perfect communication cannot be guaranteed~\cite{dasilvaSemidefiniteprogrammingbasedOptimizationQuantum2023}. Two paradigmatic noise models that frequently arise in experiments and serve as useful test cases are the dephasing and depolarizing channels: the former describes the washout of coherence properties (off-diagonal terms) of a state, while the latter represents isotropic white noise. For these reasons, we focus on these two channels as representative examples.

The dephasing channel is defined by
\begin{equation}\label{equ11}
	\Lambda_{\mathrm{deph},\nu}(\rho)=\nu\rho + (1-\nu)\mathrm{diag}(\rho),
\end{equation}
where $\mathrm{diag}(\rho)$ denotes the density matrix obtained from $\rho$ by zeroing its off-diagonal terms in the computational basis, and $\nu\in[0,1]$ denotes the visibility ($\nu=1$ corresponds to no noise).

The depolarizing channel is defined by
\begin{equation}\label{equ12}
	\Lambda_{\mathrm{depo},\nu}(\rho)=\nu\rho + (1-\nu)\mathrm{Tr}(\rho)\frac{\mathbb{I}_d}{d},
\end{equation}
with the same definition of the parameter $\nu$. For a fixed noise model and visibility $\nu$, we consider the optimal ASP with the channel instantiated by $\Lambda_{\mathrm{deph},\nu}$ or $\Lambda_{\mathrm{depo},\nu}$. Concretely, we evaluate
\begin{equation}\label{equ13}
	\beta_{n,d,\nu}^{\mathrm{chan}}
	:=\max_{\{\rho_x\},\{M_{b|y}\}}
	\frac{1}{nd^{n}}\sum_{x,y}
	\operatorname{Tr}\!\left[\Lambda_{\mathrm{chan},\nu}(\rho_x)M_{x_y|y}\right],
\end{equation}
where ``chan'' stands for either ``deph'' or ``depo''. Since the channel is explicit, we insert it directly into the objective. Nevertheless, the joint optimization over preparation states and measurements remains nonconvex and is therefore addressed using the alternating convex search method, similar to the one employed in Ref.~\cite{dasilvaSemidefiniteprogrammingbasedOptimizationQuantum2023}.

\begin{figure}
	\centering
	\includegraphics[width=8.5cm]{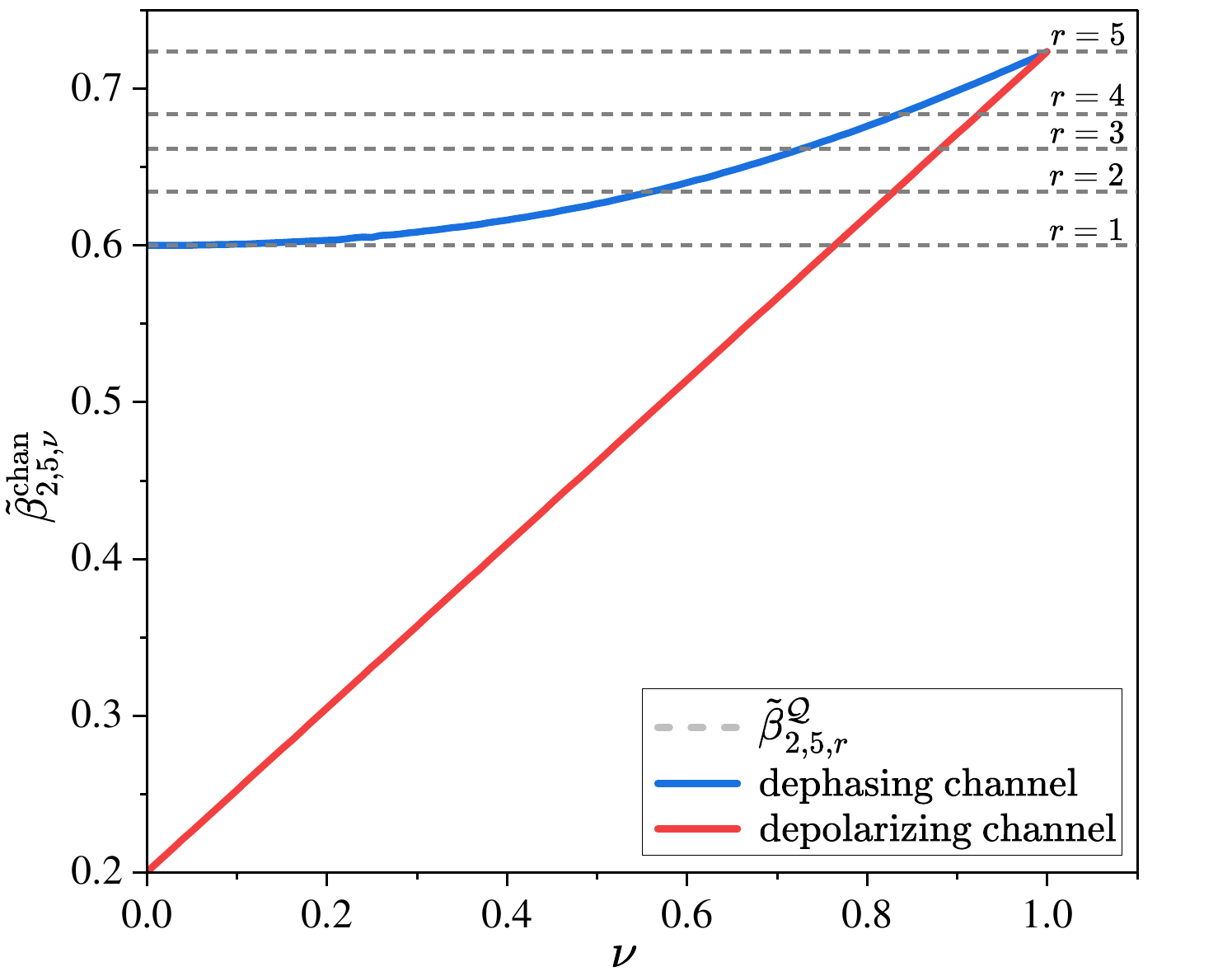}
	\caption{Numerical values $\tilde{\beta}_{2,5,\nu}^{\mathrm{chan}}$ for the optimal ASP under noisy channels as a function of the visibility $\nu$. The blue curve corresponds to the dephasing channel and the red curve to the depolarizing channel. For comparison, horizontal dashed lines indicate the previously computed Schmidt-number-dependent bounds $\tilde{\beta}_{2,5,r}^{\mathcal{Q}}$ for $r=1,\dots,5$.}
	\label{fig3}
\end{figure}

Fig.~\ref{fig3} presents the numerical values $\tilde{\beta}_{n,d,\nu}^{\mathrm{chan}}$ as a function of the visibility $\nu$ for the illustrative case $n=2$ and $d=5$. The two noise models exhibit distinct behaviors: for the dephasing channel, the optimal ASP reaches the classical bound $\beta_{2,5}^{\mathcal{L}}$ already at $\nu=0$, whereas for the depolarizing channel a significantly higher visibility is required, indicating a stronger detrimental effect on information transmission. The figure also includes the previously computed Schmidt-number-dependent bounds $\tilde{\beta}_{2,5,r}^{\mathcal{Q}}$. Comparing the visibility-dependent curves with these results allows one to directly infer the relation between channel visibility and certifiable entanglement dimensionality. In particular, for each $r$, the intersection of $\tilde{\beta}_{2,5,\nu}^{\mathrm{chan}}$ with $\tilde{\beta}_{2,5,r}^{\mathcal{Q}}$ defines a critical visibility $\nu_c(r)$, above which the channel can be certified to have an entanglement dimensionality exceeding $r$.

\section{\label{sec4:level1} Certifying the Entanglement Fidelity of Quantum Channels}

We then consider how to certify the entanglement fidelity of quantum channels in the SDI-P$\&$M scenario. Specifically, we seek to derive, solely from the observed statistics, a lower bound on the entanglement fidelity $\mathcal{F}(\Phi_\Lambda)$ of the channel. Such a bound has a clear operational meaning: it certifies that, regardless of the internal structure of the devices compatible with the data, the action of the channel cannot reduce a maximally entangled state below a certain fidelity threshold. In this sense, entanglement fidelity provides a robust and experimentally accessible figure of merit for assessing the quality of high-dimensional quantum channels.

A key ingredient of our approach is a dual SDP representation of the entanglement fidelity, which enables systematic relaxations consistent with the observed statistics. In particular, the entanglement fidelity defined in Eq.~\eqref{equ6} admits the dual formulation
\begin{equation}
	\begin{aligned}
		\mathcal{F}(\Phi_\Lambda)
		=\min_{D}\ \ & \frac{1}{d}\,\mathrm{Tr}(D) \; := f, \\
		\mathrm{s.t.}\ \ & \mathbb{I}_{\mathcal{A}} \otimes D - \Phi_\Lambda \succeq 0 ,
	\end{aligned}
	\label{equ14}
\end{equation}
where the dual variable $D$ acts on the subsystem $\mathcal{B}$. Both the primal and dual problems are strictly feasible, ensuring that strong duality holds~\cite{boydConvexOptimization2004}. A detailed derivation of this dual representation can be found in Ref.~\cite{skrzypczykSemidefiniteProgrammingQuantum2023}. To obtain computable lower bounds on $\mathcal{F}(\Phi_\Lambda)$ from observed data, we embed Eq.~\eqref{equ14} into a hierarchy of SDP relaxations based on localizing matrices~\cite{tavakoliSemidefiniteProgrammingRelaxations2024}. We begin by introducing the basic operator list
\[
\mathcal{I}
=\bigl\{
\mathbb{I}_{\mathcal{A}}\otimes\mathbb{I}_{\mathcal{B}},
\{E_{i,j}\otimes\mathbb{I}_{\mathcal{B}}\}_{i,j},
\{\rho_x\otimes\mathbb{I}_{\mathcal{B}}\}_x,
\{\mathbb{I}_{\mathcal{A}}\otimes M_{b|y}\}_{b,y}
\bigr\},
\]
where $E_{i,j}=|i\rangle\langle j|$ denotes the matrix units associated with the computational basis $\{|i\rangle\}$ used to define the maximally entangled state. We assume that the preparation states $\rho_x$ are pure, i.e., $\rho_x^2=\rho_x$, and that the measurement operators ${M_{b|y}}$ are projective, satisfying $M_{b|y}M_{b^\prime|y} = \delta_{b,b^\prime} M_{b|y}$.

For a given hierarchy level $\ell$, we define $\mathcal{O}_\ell$ as the set of all operator monomials formed by products of at most $\ell$ elements from $\mathcal{I}$. For any operator $\sigma_{\mathcal{AB}}$, we then associate the corresponding localizing matrix
\begin{equation}\label{equ15}
	\Gamma_\ell[\sigma]
	=\sum_{i,j} |i\rangle\langle j|\,
	\mathrm{Tr}\!\left[\sigma\, O_j^\dagger O_i\right],
	\  O_i,O_j\in\mathcal{O}_\ell .
\end{equation}

Under this construction, the objective function in Eq.~\eqref{equ14} reads
\(
f=\Gamma_\ell[D]_{\mathbb{I}_{\mathcal{A}}\otimes\mathbb{I}_{\mathcal{B}}}/d^2,
\)
where we adopt the shorthand $\Gamma_\ell[D] \equiv \Gamma_\ell[\mathbb{I}_{\mathcal{A}}\otimes D]$. The feasibility conditions translate into the semidefinite constraints
\[
\Gamma_\ell[\Phi_\Lambda]\succeq 0,
\ \ 
\Gamma_\ell[D]-\Gamma_\ell[\Phi_\Lambda]\succeq 0.
\]
In addition, the localizing matrix $\Gamma_\ell[\Phi_\Lambda]$ is subject to further linear constraints reflecting physical consistency with the observed data and the channel  properties. For instance, $\Gamma_\ell[\Phi_\Lambda]_{\mathbb{I}_\mathcal{A}\otimes\mathbb{I}_\mathcal{B}}=1$ for normalization, $\Gamma_\ell[\Phi_\Lambda]_{\rho_x\otimes M_{b|y}}=P(b|x,y)/d$ for compatibility with the observed statistics, and $\Gamma_\ell[\Phi_\Lambda]_{E_{i,j}\otimes \mathbb{I}_\mathcal{B}}=\delta_{i,j}/d$ for the partial-trace constraint in Eq.~\eqref{equ3}. For a complete specification of the constraints, we refer the reader to our implementation code provided in Ref.~\cite{Li2025HDChannelCertification}. We note that the above framework can be readily extended beyond the assumption of pure states. If the preparation states $\rho_x$ are mixed, they satisfy the operator inequality $\rho_x - \rho_x^2 \succeq 0$, which can be incorporated into the SDP relaxation through an additional localizing matrix $\Gamma_\ell[\rho_x - \rho_x^2]$~\cite{rochicarcellerPrepareandMeasureScenariosPhotonNumber2025}. This matrix is required to be positive semidefinite and to satisfy the corresponding physical consistency constraints, ensuring that the framework remains valid for mixed-state preparations.

Instead of enforcing compatibility with the full set of observed statistics, one may adopt a witness based certification approach, namely by imposing the observed ASP $\alpha_{n,d}$ as a constraint. In this case, the SDP relaxation can be written as
\begin{equation}\label{equ16}
	\begin{aligned}
		\underset{\Gamma_\ell[D],\,\Gamma_\ell[\Phi_\Lambda]}{{\min}}
		& \ \  \frac{1}{d^2}\Gamma_\ell[D]_{\mathbb{I}_{\mathcal{A}}\otimes\mathbb{I}_{\mathcal{B}}} \\
		\mathrm{s.t.~~~~}
		& \ \  \Gamma_\ell[\Phi_\Lambda] \succeq 0, \\
		& \ \  \Gamma_\ell[D] - \Gamma_\ell[\Phi_\Lambda] \succeq 0, \\
		& \ \  \sum_k \mathrm{Tr}\!\left(C_k\,\Gamma_\ell[\Phi_\Lambda]\right) = \alpha_{n,d},
	\end{aligned}
\end{equation}
together with additional linear constraints discussed above. Here, $C_k$ denote the coefficient matrices defining the ASP witness.  Solving the resulting SDP at increasing hierarchy levels yields a sequence of progressively tighter lower bounds on the entanglement fidelity $\mathcal{F}(\Phi_\Lambda)$ that are compatible with the observed value of $\alpha_{n,d}$.

Building on the analysis in the preceding section, we can determine the certified entanglement fidelity compatible with the observed value of $\alpha_{n,d}$ in some specific instances. Specifically, when $\alpha_{n,d} = \beta_{n,d}^{\mathcal{L}}$, the channel $\Lambda$ may, in the worst case, be entanglement breaking. In this situation, the associated Choi state $\Phi_\Lambda$ is separable, and its maximal achievable fidelity with a maximally entangled state is bounded by $1/d$. Furthermore, when $\alpha_{n,d} = \beta_{n,d}^{\mathcal{Q}}$, the channel $\Lambda$ must necessarily be nonentanglement breaking; correspondingly, $\Phi_\Lambda$ is equivalent to a maximally entangled state up to local unitary transformations, and the entanglement fidelity can attain its maximal value of unity.

\begin{figure}
	\centering
	\includegraphics[width=8.1cm]{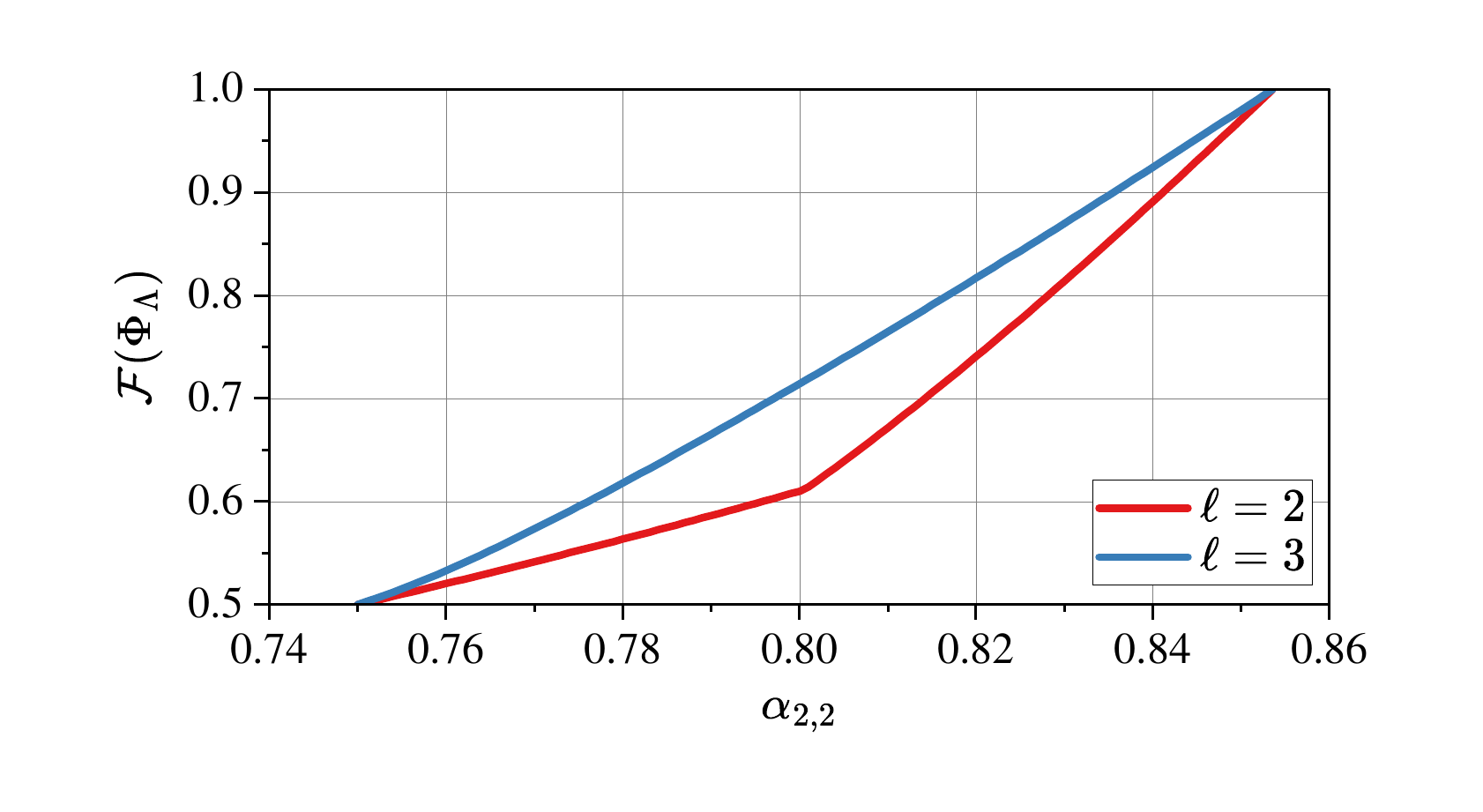}
	\caption{ Numerical lower bounds on the entanglement fidelity $\mathcal{F}(\Phi_\Lambda)$ are shown as a function of the observed ASP $\alpha_{2,2}$, with $\alpha_{2,2} \in [\beta_{2,2}^{\mathcal{L}}, \beta_{2,2}^{\mathcal{Q}}]$. The red and blue curves correspond to hierarchy levels $\ell=2$ and $\ell=3$, respectively, with the latter already close to convergence and providing a tight lower bound on the entanglement fidelity.}
	\label{fig4}
\end{figure}

As a concrete example, we focus on the minimal nontrivial case $n=d=2$, which already suffices to illustrate the certified lower bound on the entanglement fidelity  $\mathcal{F}(\Phi_\Lambda)$. The corresponding numerical results are displayed in Fig.~\ref{fig4}. As anticipated, the numerical bounds lie between the two values $1/2$ and $1$, in full agreement with the above analysis. Notably, the entanglement fidelity varies continuously with the observed statistics, in contrast to entanglement dimensionality, which exhibits intrinsically threshold-like behavior.

\section{\label{sec5:level1}CONCLUSION}

In this work, we have developed a SDI framework for certifying high-dimensional quantum channels based solely on observed P$\&$M statistics. By explicitly incorporating the full set of structural constraints intrinsic to Choi states, our approach has fully exploited the CJ isomorphism to characterize quantum channels in a physically consistent manner. Within this framework, we have constructed systematic certification schemes for the entanglement dimensionality and the entanglement fidelity of quantum channels, respectively. 

The entanglement dimensionality of quantum channels has been certified using a tailored witness, whose Schmidt-number-dependent bounds have been obtained through numerical methods. Investigations into dephasing and depolarizing noise channels have further demonstrated the practical utility of our approach, providing insight into the relationship between noise parameters and the certifiable entanglement dimensionality. Beyond dimensionality certification, we have developed a systematic method for certifying entanglement fidelity based on a hierarchy of SDP relaxations, yielding lower bounds that are compatible with either the full set of observed statistics or an experimentally accessible witness value.

Existing studies have primarily focused on certifying the entanglement dimensionality of high-dimensional quantum channels, typically relying on fully trusted internal devices. In contrast, our work certifies both entanglement dimensionality and entanglement fidelity without assuming trusted devices. The entanglement fidelity is crucial because it quantifies the actual strength of entanglement preserved by a channel, which can vary significantly even among channels with the same entanglement dimensionality. By combining dimensionality and fidelity certifications, we provide a more complete assessment of channel performance and enhance its applicability to practical quantum communication tasks. Nevertheless, we observe that certifying entanglement fidelity often requires relatively high levels in the SDP hierarchy to achieve convergence, whereas lower levels are sufficient for other certification tasks~\cite{pauwelsAlmostQuditsPrepareandmeasure2022,2019wangCharacterisingCorrelations,2025jiaCharacterizingSet}. This discrepancy may stem from certain constraints in the SDP hierarchy not being adequately considered, or from the inherent complexity of the entanglement fidelity certification. 

\section*{Acknowledgment}
This work is supported by the National Natural Science Foundation of China (Grants No. 62571060, No. 62171056, and No. 62220106012).

\appendix

\section{\label{app} ALTERNATING CONVEX SEARCH METHOD}

We solve Eq.~\eqref{equ9} using the alternating convex search method that alternates between optimizing $\{\rho_x\}$, $\{M_{b|y}\}$, and $\Phi_\Lambda$. Each subproblem is convex when the other blocks are fixed, and the objective increases monotonically. The only nontrivial component is the treatment of $\Phi_\Lambda$. Generating suitable random initial operators that respect a target Schmidt number bound, as well as relaxing the constraint $\mathrm{SN}(\Phi_\Lambda)\le r$, requires dedicated techniques. These are described in Secs.~\ref{app:subsec2} and \ref{app:subsec3}. 

\subsection{\label{app:subsec2} Random initialization of $\Phi_\Lambda$}

The optimization procedure begins with randomized initial values for all optimization variables, namely the ensembles $\{\rho_x\}$, the POVMs $\{M_{b|y}\}$, and the Choi operator $\Phi_\Lambda$. While the random generation of quantum states and measurement operators can be carried out straightforwardly using standard techniques such as sampling random pure states or Haar-random unitaries, the initialization of $\Phi_\Lambda$ is substantially more delicate. In particular, producing Choi states that satisfy a prescribed Schmidt number bound cannot be achieved by naive random sampling and thus requires a dedicated construction. 
Below we describe the construction used in our numerics.

\subsubsection{\label{app:subsubsec2-1} Concept and guarantees}  
We generate a random CPTP map $\Lambda$ by specifying a finite Kraus decomposition
\[
\Lambda(\rho)=\sum_{k=1}^{K} A_k\,\rho\,A_k^\dagger,\qquad
\sum_{k=1}^{K} A_k^\dagger A_k=\mathbb{I}_d,
\]
and then form the corresponding Choi state $\Phi_\Lambda$. If every Kraus operator $A_k$ in the chosen decomposition has $\operatorname{rank}(A_k)\le r$, then $\Phi_\Lambda$ necessarily has operator Schmidt number $\mathrm{SN}(\Phi_\Lambda)\le r$ \cite{chruscinskiPartiallyEntanglementBreaking2006}. Thus the construction below yields random Choi states compatible with the desired Schmidt number bound (and, when arranged explicitly, can be used to produce examples with effectively exact Kraus ranks equal to $r$ up to numerical precision).

\subsubsection{\label{app:subsubsec2-2} Algorithmic construction}  
Fix the system dimension $d$, target Schmidt number parameter $r\le d$, and choose the number of Kraus operators $K\ge1$. The random channel initialization proceeds in three main steps:

\begin{enumerate}
	\item[1)] \emph{Generate $\,K$ raw Kraus matrices of rank $\le r$.} For each $k=1,\dots,K$ form a complex $d\times d$ matrix $A_k^{(0)}$ whose column (or singular) span has dimension $\le r$. Two convenient constructions are:
	\begin{itemize}
		\item \emph{Exact-rank variant:} draw independent Haar-random unitaries $U,V\in\mathrm{U}(d)$, take the first $r$ columns $U_{: ,1:r}$ and $V_{: ,1:r}$, choose a diagonal matrix $S\in\mathbb{C}^{r}$ with nonzero entries, and set
		\[
		A_k^{(0)} = U_{:,1:r}\, S\, V_{:,1:r}^\dagger,
		\]
		which has rank exactly $r$ (up to numerical tolerance).
		\item \emph{Rank-$\le r$ variant:} draw random complex $d\times r$ and $r\times d$ matrices $X$ and $Y$ with i.i.d.\ Gaussian entries and set
		\[
		A_k^{(0)} = X Y,
		\]
		which has rank at most $r$.
	\end{itemize}
	
	\item[2)] \emph{Normalize to enforce the trace-preserving condition.} Compute
	\[
	T=\sum_{k=1}^K (A_k^{(0)})^\dagger A_k^{(0)}.
	\]
	Provided $T$ is invertible on its support, form the inverse square root $T^{-1/2}$ (regularizing small eigenvalues as needed for numerical stability) and update
	\[
	A_k := A_k^{(0)}\, T^{-1/2},\qquad k=1,\dots,K,
	\]
	so that $\sum_k A_k^\dagger A_k = \mathbb{I}_d$ holds up to numerical precision. This step maps the raw collection $\{A_k^{(0)}\}$ to a valid set of Kraus operators for a CPTP map.
	
	\item[3)] \emph{Form the Choi state.} With the normalized Kraus operators $\{A_k\}$ and the maximally entangled vector $|\Phi_d^+\rangle$, compute
	\[
	\Phi_\Lambda=\sum_{k=1}^K (\mathbb{I}_d\otimes A_k)\,|\Phi_d^+\rangle\langle\Phi_d^+|\,(\mathbb{I}_d\otimes A_k)^\dagger.
	\]
	Enforce Hermiticity and exact unit trace by setting $\Phi_\Lambda\leftarrow \tfrac{1}{2}(\Phi_\Lambda+\Phi_\Lambda^\dagger)$ and $\Phi_\Lambda\leftarrow \Phi_\Lambda/\operatorname{Tr}(\Phi_\Lambda)$ if numerical drift is present.
\end{enumerate}

\subsubsection{\label{app:subsubsec2-3} Practical remarks and checks}
The construction guarantees $\Phi_\Lambda\succeq0$, $\operatorname{Tr}(\Phi_\Lambda)=1$ and $\operatorname{Tr}_\mathcal{B}(\Phi_\Lambda)=\mathbb{I}_d/d$ up to negligible numerical drift; moreover, if every Kraus operator $A_k$ is drawn with $\operatorname{rank}(A_k)\le r$ then $\mathrm{SN}(\Phi_\Lambda)\le r$ by construction. For instances intended to realize $\mathrm{SN}(\Phi_\Lambda)=r$ one should use the exact-rank variant with nondegenerate singular values and check the numerical ranks of the matrices $A_k$. For numerical stability, form the inverse square root $T^{-1/2}$ of $T=\sum_k (A_k^{(0)})^\dagger A_k^{(0)}$ with small-eigenvalue regularization (e.g.\ threshold $ \sim 10^{-14}$ or problem-dependent tolerance), and afterwards check that $\sum_k A_k^\dagger A_k \approx \mathbb{I}_d$. 

\subsection{\label{app:subsec3} Relaxing the Constraint $\mathrm{SN}(\Phi_\Lambda)\le r$}

\subsubsection{\label{app:subsubsec3-1} Generalized Doherty-Parrilo-Spedalieri hierarchy}

Let $\mathcal{S}$ denote the set of separable bipartite quantum states. A standard convex optimization approach to approximate $\mathcal{S}$ relies on the DPS hierarchy~\cite{2002dohertyDistinguishingSeparable}. This construction defines a nested sequence of convex sets,
\[
\mathcal{S}^1 \supseteq \mathcal{S}^2 \supseteq \cdots \supseteq \mathcal{S},
\]
where $\mathcal{S}^k$ consists of bipartite states admitting a positive-partial-transpose (PPT) $k$-symmetric extension (see, e.g., Ref.~\cite{2002dohertyDistinguishingSeparable} for precise definitions). Membership testing of the form $\sigma_{\mathcal{AB}} \in \mathcal{S}^k$ can be formulated as a SDP problem, and the hierarchy is complete in the sense that $\mathcal{S}^k$ converges to $\mathcal{S}$ in the limit $k \to \infty$.

To adapt this framework to states with Schmidt number, one constructs a generalized DPS hierarchy~\cite{2020weilenmannEntanglementDetection}. Formally, a bipartite state $\sigma_{\mathcal{AB}}$ belongs to $\mathcal{S}_r^k$ if there exists a positive semidefinite operator $\mathcal{U}$ acting on $\mathcal{A}\mathcal{A}'\mathcal{B}'\mathcal{B}$ such that
\[
\Pi_r^\dagger \, \mathcal{U} \, \Pi_r = \sigma_{\mathcal{AB}}, \qquad
\frac{\mathcal{U}}{r} \in \mathcal{S}^k, \qquad
\mathrm{Tr}(\mathcal{U}) = r,
\]
where $\Pi_r := \mathbb{I}_{\mathcal{A}} \otimes \sqrt{r}\,|\Phi_r^{+}\rangle_{\mathcal{A}'\mathcal{B}'} \otimes \mathbb{I}_{\mathcal{B}}$, and $\dim \mathcal{A}' = \dim \mathcal{B}' = r$. The separability condition for $\mathcal{U}/r$ is imposed with respect to the bipartition $\mathcal{A}\mathcal{A}' \,|\, \mathcal{B}'\mathcal{B}$. For $r=1$, this definition reduces to the standard DPS hierarchy.

By construction, the resulting sets satisfy
\[
\mathcal{S}_r^1 \supseteq \mathcal{S}_r^2 \supseteq \cdots \supseteq \mathcal{S}_r,
\]
each membership test $\sigma_{\mathcal{AB}} \in \mathcal{S}_r^k$ admits a SDP formulation, and increasing $k$ yields progressively tighter outer approximations that converge to $\mathcal{S}_r$ in the asymptotic limit (see Ref.~\cite{2020weilenmannEntanglementDetection} for formal statements and proofs).

The generalized DPS hierarchy offers high discrimination power but is computationally demanding: both the matrix dimensions and the number of SDP constraints grow rapidly with $k$ and with the local dimensions. In numerical implementations one therefore typically employs small $k$ (often $k=1,2$) and treats the resulting test as a relatively tight (but not exact) certificate for membership in $\mathcal{S}_r$.

\subsubsection{\label{app:subsubsec3-2} Generalized reduction map}

As a computationally lightweight alternative, we also employ a criterion based on the generalized reduction map~\cite{terhalSchmidtNumberDensity2000}, defined as
\[
\Lambda_R(\rho) := \mathrm{Tr}(\rho)\,\mathbb{I}_d - \frac{1}{r}\,\rho .
\]
Applying this map locally to subsystem $\mathcal{B}$ of a bipartite state $\sigma_{\mathcal{AB}}$ yields the operator inequality
\begin{equation}
	(\mathrm{id} \otimes \Lambda_R)(\sigma_{\mathcal{AB}})
	= \mathrm{Tr}_{\mathcal{B}}(\sigma_{\mathcal{AB}}) \otimes \mathbb{I}_{\mathcal{B}} - \frac{1}{r}\,\sigma_{\mathcal{AB}} \succeq 0 ,
	\label{eq:redcrit}
\end{equation}
which is satisfied by all states with Schmidt number at most $r$~\cite{terhalSchmidtNumberDensity2000}. Enforcing Eq.~\eqref{eq:redcrit} therefore yields an efficiently computable outer approximation to the set $\mathcal{S}_r$, providing a necessary condition that is particularly well suited to iterative optimization procedures. Although this criterion is generally weaker than the generalized DPS hierarchy, it has proven effective for Schmidt number certification in related contexts~\cite{terhalSchmidtNumberDensity2000,2023wyderkaConstructionEfficient,mallickHigherdimensionalentanglementDetectionQuantumchannel2025}.

\end{document}